\documentclass[copyright,creativecommons]{eptcs}
\usepackage{float} 

\providecommand{\event}{HCVS 2016} 
\usepackage{breakurl}             

\usepackage[dvips]{graphicx}
\usepackage{latexsym}
\usepackage{mdframed}
\usepackage{rotating}
\usepackage{listings}

\usepackage{amsmath,amssymb,amsfonts}

\usepackage{amsthm} 
\usepackage[usenames,dvipsnames,svgnames,table]{xcolor}
\usepackage{color}
\usepackage{fancybox}
\usepackage{xspace}
\usepackage{ar-alg}
\usepackage[inference]{semantic}
\usepackage{pgf,tikz}
\usetikzlibrary{arrows,automata}
\usepackage{rotating}
\usepackage{cite}
\usepackage{frameit} 

\usepackage{mathtools,wrapfig}
\usepackage{alltt,enumitem,comment,hhline}
\usepackage{floatflt}
\usepackage{multicol}
\usepackage{multirow}
\usepackage{appendix}

\usepackage{ezmath}
\usepackage{ezcode}
\usepackage{ezcodeBarn}
\usepackage{flushend}

\newcommand{\init}{\mathit{init}}

\newcommand{\Vars}{v}

\newcommand{\ctlSymb}{\psi}

\newcommand{\ctlsat}{\models_{\mathit{CTL}}}

\newcommand{\qsymWF}{\mathit{wf}}

\newcommand{\Inv}{\mathit{inv}}

\newcommand{\limp}{\rightarrow}

\newcommand{\ltrue}{\mathit{true}}

\newcommand{\loc}{\ell}

\newcommand{\pc}{\mathit{pc}}
\newcommand{\p}{\mathtt{p}}

\newcommand{\Tool}{\textsc{E-HSF}\xspace}

\newcommand{\makeFramedRule}[3]{
  \begin{mdframed}
    \vspace{-0.6cm}
    \mbox{}\\
    \flushleft
    #1
    \begin{equation*}
      \begin{array}[t]{c}
        #2 
      \end{array}
    \end{equation*}
    \centering
    \rule{.7\linewidth}{.5pt}\\ 
    \ensuremath{#3}
    \vspace{0.2cm}
  \end{mdframed}
}

\newcommand{\makeFramedRuleSS}[3]{
  \begin{mdframed}
    \vspace{-0.6cm}
    \mbox{}\\
    \flushleft
    #1
    \begin{equation*}
      \begin{array}[t]{rcl}
        #2 
      \end{array}
    \end{equation*}
    \centering
    \rule{.7\linewidth}{.5pt}\\ 
    \ensuremath{#3}
    \vspace{0.2cm}
  \end{mdframed}
}

\newcommand{\skolemRelSymbol}{\mathit{rel}}

\newcommand{\algEHSF}{\textsc{E-HSF}\xspace}

\newcommand{\funTemplateOf}[2]{\textsc{Templ}(#1)(#2)}

\newcommand{\cc}[1]{\multicolumn{2}{|c|}{#1}}

\newcommand{\ipYes}{\checkmark\xspace}
\newcommand{\ipNo}{$\times$}

\newcommand{\inv}{\mathit{inv}}

\newcommand{\relnext}{\mathit{next}}

{
\setlength{\fboxsep}{6pt}%
\begin{Sbox}\begin{minipage}}%
{\end{minipage}\end{Sbox}\fbox{\TheSbox}
}

\newtheorem{theorem}{Theorem}

\newcommand{\nextRel}{\mathit{next}}

\newcommand{\computations}[2]{\ensuremath{\Pi_{#1}(#2)}}
\newcommand{\program}{\ensuremath{P}}
\newcommand{\sat}[3]{\ensuremath{#1, #2 \models #3}}

\newcommand{\modelsT}{\models_{\mathcal{T}}}

\newcommand{\pathE}{E} 
\newcommand{\pathA}{A} 
\newcommand{\ltlNext}{X} 
\newcommand{\ltlG}{G} 
\newcommand{\ltlF}{F} 
\newcommand{\ltlU}{U}  

\newcommand{\rank}{\mathit{rank}}

\usepackage{capt-of}

\newcommand{\ruleCTLInit}{\textsc{RuleCtlInit}\xspace}

\newcommand{\ruleCTLDecompUni}{\textsc{RuleCtlDecompUni}\xspace}
\newcommand{\ruleCTLDecompBin}{\textsc{RuleCtlDecompBin}\xspace}

\newcommand{\ruleCTLEX}{\textsc{RuleCtlEX}\xspace}
\newcommand{\ruleCTLEG}{\textsc{RuleCtlEG}\xspace}
\newcommand{\ruleCTLEU}{\textsc{RuleCtlEU}\xspace}

\newcommand{\ruleCTLAX}{\textsc{RuleCtlAX}\xspace}
\newcommand{\ruleCTLAG}{\textsc{RuleCtlAG}\xspace}
\newcommand{\ruleCTLAU}{\textsc{RuleCtlAU}\xspace}

\title{Efficient CTL Verification via Horn Constraints Solving}
\author{Tewodros A{.} Beyene
\institute{fortiss GmbH \\ Munich, Germany} 
\email{beyene@fortiss.org}
\and Corneliu Popeea
\institute{CQSE GmbH\\ Munich, Germany} 
\email{popeea@cqse.eu}
\and Andrey Rybalchenko
\institute{Microsoft Research\\ Cambridge, UK}
\email{rybal@microsoft.com}
}
\def\titlerunning{Efficient CTL Verification via Generic Horn Constraints Solving}
\def\authorrunning{T.A. Beyene, C. Popeea \& A. Rybalchenko}

\begin{document}
\maketitle
\begin{abstract} 
The use of temporal logics has long been recognised as a fundamental approach to
the formal specification and verification of reactive systems. 
In this paper, we take on the problem of automatically verifying a
temporal property, given by a CTL formula, for a given (possibly
infinite-state) program. 
We propose a method based on encoding the problem as a set of Horn
constraints.  
The method takes a program, modeled as a transition system, and a
property given by a CTL formula as input. 
It first generates a set of forall-exists quantified Horn
constraints and well-foundedness constraints by exploiting the
syntactic structure of the CTL formula. 
Then, the generated set of constraints are solved by applying an
off-the-shelf Horn constraints solving engine. 
The program is said to satisfy the property if and only if
the generated set of constraints has a solution. 
We demonstrate the practical promises of the method by applying it on
a set of challenging examples. 
Although our method is based on a generic Horn constraint solving
engine, it is able to outperform state-of-art methods specialised
for CTL verification. 
\end{abstract}


\section{Introduction}
\label{sec-ctl-intro}

Since Pnueli's pioneering work~\cite{Pnueli1977}, the use of temporal logics has
long been recognised as a fundamental approach to the formal
specification and verification of reactive systems~\cite{Manna1992,
  Emerson1991TM}.  
Temporal logics allow precise specification of complex properties.
There have been decades of effort on temporal verification of finite state
systems~\cite{Kupferman2000, Burch1990, Clarke02treelikeCEX,
  Clarke1983AVF}. 
For CTL and other state-based properties, the standard procedure is to adapt
“bottom-up” (or “tableaux”) techniques for reasoning on finite-state
systems.
In addition, various classes of temporal logics support
model-checking whose success over the last twenty years allows
large and complex (finite) systems to be verified automatically~\cite{Burch1990,
  Clarke1990TLM, McMillan1993SMC, Hassan2012IIC}.
In recent decades, however, the research focus has shifted to
infinite-state systems in general and to software systems in
particular as ensuring correctness for software is
in high demand. 
Most algorithms for verifying CTL properties on infinite-state systems
typically involve first abstracting the state space into a finite-state
model, and then applying finite reasoning strategies on the abstract
model. 
There is also a lot of effort on algorithms that are focused on a
particular fragment of CTL, such as the universal
fragment~\cite{Penczek2002BMC} and the existential
fragment~\cite{GurfinkelWC06}, or some
particular classes of infinite-state systems such as pushdown
processes \cite{Song2011, Song2013, Walukiewicz2000,
  walukiewicz2001pushdown} or parameterised
systems~\cite{Emerson1996, Demri2010checkingctl}.

In this paper, we take on the problem of automatically verifying CTL
properties for a given (possibly infinite-state) program.  
We propose a method based on solving a set of forall-exists quantified
Horn constraints.
Our method takes a program $\program$ modeled by a transition system
$(\init(v), \relnext(v,v'))$ and a property given by a CTL formula
$\varphi(v)$, and then it checks if $\program$ satisfies $\varphi(v)$,
i.e., if $(\init(v), \relnext(v, v'))\ctlsat \varphi(v)$. 
The method first generates a set of forall-exists quantified Horn constraints
with well-foundedness conditions by exploiting the syntactic structure
of the CTL formula $\varphi(v)$.
It then solves the generated set of Horn constraints by applying an
off-the-shelf solving engine \algEHSF~\cite{ehsf} for such
constraints.
We claim that $\program$ satisfies $\varphi(v)$ if and only if
the generated set of Horn constraints has a solution. 
We demonstrate the practical applicability of the method by presenting
experimental evaluation using examples from the PostgreSQL database
server, the SoftUpdates patch system, the Windows OS kernel.

The rest of the paper is organised as follows. 
We start by summarising the syntax and semantics of CTL and by giving a brief
introduction to forall-exists quantified Horn constraints and their
solver \algEHSF in Section~\ref{sec:prelims}.
In Section~\ref{sec:proof-system}, we present our CTL proof system
that generates a set of forall-exists quantified Horn constraints for
a given verification problem.
We illustrate application of the proof rules on an example in
Section~\ref{sec:cons-gen}.
The experimental evaluation of our method is given in
Section~\ref{sec:eval}.
Finally, we present a brief discussion on related work in
Section~\ref{sec:rel-work} and concluding remarks in
Section~\ref{sec:concl}.


\section{Preliminaries} 
\label{sec:prelims}
\subsection{CTL basics} 

In this section, we review the syntax and the semantics of the logic
CTL following~\cite{KestenTCS95}. 
Let $\mathcal{T}$ be a first order theory and $\modelsT$ denote its
satisfaction relation that we use to describe sets and relations
over program states.
Let $c$ range over assertions in $\mathcal{T}$.
A CTL formula $\varphi$ is defined by the following grammar using
the notion of a path formula~$\phi$.
\begin{equation*}
  \begin{array}[t]{@{}r@{\;::=\;}l@{}}
  \varphi &
    c \mid
    \varphi \land \varphi  \mid
    \varphi \lor \varphi   \mid
    \pathA \, \phi \mid
    \pathE \, \phi \\ [\jot]
    \phi & \ltlNext \varphi \mid \ltlG \varphi \mid \varphi \ltlU \varphi
  \end{array}
\end{equation*}
$\ltlNext$, $\ltlG$, and $\ltlU$ are called temporal operators, and
$\pathA$ and $\pathE$ are called path quantifiers.  
A CTL formula whose principal operators are a pair QT, where Q is a
path quantifier and T is a temporal operator, and which does not 
contain any additional temporal operators or path quantifiers is
called a basic CTL formula.
As usual, we define  $\ltlF \varphi = (\ltrue~ \ltlU \varphi)$.
The satisfaction relation $\program\models \varphi$ holds if and only
if for each $s$ such that $\init(s)$ we
have~$\sat{\program}{s}{\varphi}$.
We define $\sat{\program}{s}{\varphi}$ as follows using an auxiliary
satisfaction relation $\sat{\program}{\pi}{\phi}$.

\begin{equation*}
  \begin{array}[t]{@{}l@{\text{ iff }}l@{}}
    \sat{\program}{s}{c} 
    &
    s \modelsT c \\[\jot]
    \sat{\program}{s}{\varphi_1 \land \varphi_2} 
    &
    \sat{\program}{s}{\varphi_1} \text{ and }
    \sat{\program}{s}{\varphi_2}\\[\jot]
    \sat{\program}{s}{\varphi_1 \lor \varphi_2} 
    &
    \sat{\program}{s}{\varphi_1} \text{ or }
    \sat{\program}{s}{\varphi_2}\\[\jot]
    \sat{\program}{s}{\pathA\, \phi} 
    &
    \text{for all $\pi \in \computations{\program}{s}$ holds } 
    \sat{\program}{\pi}{\phi} \\[\jot]
    \sat{\program}{s}{\pathE\, \phi}
    &
    \text{exists $\pi \in \computations{\program}{s}$ such that }
    \sat{\program}{\pi}{\phi} \\[\jot]
    \sat{\program}{\pi}{\ltlNext \varphi}
    &
    \pi = s_1, s_2, \ldots \text{ and }
    \sat{\program}{s_2}{\varphi} \\[\jot]
    \sat{\program}{\pi}{\ltlG \varphi}
    &
    \pi = s_1, s_2, \ldots \text{for all $i\geq 1$ holds }
    \sat{\program}{s_i}{\varphi} \\[\jot]
    \sat{\program}{\pi}{\varphi_1 \ltlU \varphi_2}
    &
    \begin{array}[t]{@{}l@{}}
      \pi = s_1, s_2, \ldots 
      \text{ and exists $j\geq 1$ such that }\\[\jot]
      \sat{\program}{s_j}{\varphi_2} \text{ and }
      \sat{\program}{s_i}{\varphi_1} \text{ for } 1 \leq i < j
    \end{array}
  \end{array}
\end{equation*}
In this paper, we represent a satisfaction relation $\program\models
\varphi$ by the relation $\program\ctlsat \varphi$ to
explicitly indicate that $\varphi$ is a CTL formula.
We call such relation a CTL satisfaction, and $\varphi$ is said to be
its formula.

\subsection{The solving algorithm \algEHSF} 
\label{subsec-ehsf}

Our proof rules are automated using the \algEHSF engine for resolving
forall-exists Horn-like clauses extended with well-foundedness
criteria.

We skip the syntax and semantics of the clauses targeted by this
system --- see \cite{ehsf} for more details. 
Instead, we illustrate these clauses with the following example:
\begin{equation*}
    \begin{array}[t]{@{}l@{\qquad}l@{}}
      x \geq 0 \limp \exists y: x \geq y \land \mathit{rank}(x, y), &
      \mathit{rank}(x, y) \limp \mathit{ti}(x, y),\\[\jot]
      \mathit{ti}(x, y) \land \mathit{rank}(y, z) \limp \mathit{ti}(x,
      z), &
      \mathit{dwf}(ti).
\end{array}
\end{equation*}

Intuitively, these clauses represent an assertion over the interpretation of
``query symbols'' $\mathit{rank}$ and~$\mathit{ti}$ (the predicate
$\mathit{dwf}$ represents disjunctive well-foundedness, and is not a
query symbol).
The semantics of these clauses maps each predicate symbol 
occurring in them into a constraint over~$v$.
Specifically, the above set of clauses has a solution that maps both
$\mathit{rank}(x, y)$ and $\mathit{ti}(x, y)$ to the  constraint  $(x
\geq 0 \land y \geq x-1)$.

\algEHSF resolves clauses like the above using  a CEGAR scheme to
discover witnesses for existentially quantified variables.  
The refinement loop collects a global constraint that declaratively
determines which witnesses can be chosen. 
The chosen witnesses are used to replace existential quantification,
and then the resulting universally quantified clauses are passed to a
solver for such clauses. 
At this step, we can benefit from emergent tools in the area of
solving Horn clauses over decidable theories, e.g.,
HSF~\cite{GrebenshchikovTACAS12} or $\mu$Z~\cite{muz}.
Such a solver either finds a solution, i.e., a model for uninterpreted
relations constrained by the clauses, or returns a counterexample,
which is a resolution tree (or DAG) representing a contradiction. 
\algEHSF turns the counterexample into an additional constraint on the
set of witness candidates, and continues with the next iteration of
the refinement loop. 
Notably, this refinement loop conjoins constraints that are obtained
for all discovered counterexamples. 
This way \algEHSF guarantees that previously handled counterexamples
are not rediscovered and that a wrong choice of witnesses can be mended.

For the existential clause above, \algEHSF introduces a
witness/Skolem relation $\skolemRelSymbol$ over variables $x$ and
$y$, i.e., $x\geq0 \land \skolemRelSymbol(x,y) \limp x \geq y \land
\mathit{rank}(x, y)$.
For each $x$ such that $x\geq0$, we require the skolem relation to
provide the corresponding value for $y$, i.e., we require all such
$x$ is in the domain of the Skolem relation. 
This is encoded by an additional clause $x\geq0 \limp \exists y:
\skolemRelSymbol(x,y)$. 
In the \algEHSF approach, the search space of a skolem relation 
$\skolemRelSymbol(x,y)$ is restricted by a template function
$\funTemplateOf{\skolemRelSymbol}{x,y}$.
In general, \algEHSF requires such template functions to be given by
the user. 


\section{Proof system}
\label{sec:proof-system}

Our CTL verification method encodes the verification problem as a problem of
solving forall-exists quantified Horn constraints with well-foundedness
conditions. 
This is done by applying a proof system that consists of various proof
rules for handling different kinds of CTL formulas.
This proof system is based on a deductive proof system for CTL*
from~\cite{KestenTCS95} which is adapted in this work to be suitable 
from the perspective of constraint generation for a CTL satisfaction.

Given a transition system $(\init(v),\relnext(v,v'))$ and a CTL
formula $\varphi(v)$, the appropriate proof rules are used from the
proof system to generate the corresponding set of Horn constraints for
the CTL satisfaction $(\init(v), \relnext(v, v'))\ctlsat \varphi(v)$.
There are two sets of proof rules in the proof system.  

\subsection{Proof rules for decomposition}

These proof rules are applied recursively to a CTL satisfaction whose
formula is neither an assertion nor a basic CTL formula.  
The proof rules decompose the given CTL formula into new
sub-formulas by following the nesting structure of the formula.
Then, the original satisfaction is reduced to new satisfactions over the
new sub-formulas and a Horn constraint relating the new
satisfactions.

There are different proof rules depending on the outermost operator
of the formula. 
One case is when the given formula $f(\psi(v))$ nests another
formula $\psi(v)$ such that the outermost operator $f$ is a pair of a
temporal path operator and a unary temporal state operator, i.e., $f
\in \{AX, AG, EX, EG\}$.
The corresponding proof rule $\ruleCTLDecompUni$ is given in
Figure~\ref{fig-ctl-proof-rule-decompUni} that shows how such satisfactions
are decomposed.
\begin{figure}[h]
  \makeFramedRule{Find an assertion $q(v)$ such that:}{ 
    (p(v), \relnext(v, v')) \ctlsat f(q(\Vars)) \quad (q(v), \relnext(v,
    v')) \ctlsat \ctlSymb(\Vars)}{ 
    (p(v), \relnext(v, v'))\ctlsat f(\ctlSymb(\Vars))}  
  \caption{Proof rule \ruleCTLDecompUni}
  \label{fig-ctl-proof-rule-decompUni}
\end{figure}
Another case is when the given formula has a structure $f(\psi_1(v),\
\psi_2(v))$ nesting the formulas $\psi_1(v)$ and $\psi_2(v)$ such that
the outermost operator $f$ is either a pair of a temporal path
operator and the state operator until or a disjunction/conjunction, i.e., $f
\in \{AU, EU, \land, \lor\}$.
Note that when $f$ is $\land$ (resp. $\lor$), the given formula
$f(\psi_1(v),\ \psi_2(v))$ corresponds to $\psi_1(v)\land
\psi_2(v)$ (resp $\psi_1(v) \lor \psi_2(v)$).
The corresponding proof rule $\ruleCTLDecompBin$ is given in
Figure~\ref{fig-ctl-proof-rule-decompBin} that shows how such satisfactions
are decomposed.
\begin{figure}[h]
  \makeFramedRule{Find assertions $q_1(v)$ and $q_2(v)$
    such that:}{
    p(v) \limp f(q_1(v), q_2(v)), \\
    (q_1(v), \relnext(v, v')) \ctlsat \ctlSymb_1(v)  \quad (q_2(v),
    \relnext(v, v')) \ctlsat \ctlSymb_2(v)}{
    (p(v), \relnext(v, v'))\ctlsat f(\ctlSymb_1(\Vars),
    \ctlSymb_2(\Vars))
  }
  \caption {Proof rule \ruleCTLDecompBin}
  \label{fig-ctl-proof-rule-decompBin}
\end{figure}

\subsection{Proof rules for constraints generation}

This set of proof rules is applied to a CTL satisfaction whose formula
is either an assertion or a basic CTL formula.
Any CTL satisfaction can be decomposed into a set of such simple CTL
satisfactions by applying the proof rules from the previous section. 
The next step will be to generate forall-exists quantified Horn constraints
(possibly with well-foundedness condition) that constrain a set of
auxiliary assertions over program states. 

The simplest of all is the proof rule \ruleCTLInit, see
Figure~\ref{fig-ctl-proof-rule-init}, which is applied when the CTL
formula is an assertion. \\
\begin{minipage}{.48\textwidth} 
\begin{figure}[H]
  \makeFramedRule{}{
    p(v) \limp \ctlSymb(\Vars)
  }{
    (p(v), \relnext(v, v')) \ctlsat \ctlSymb(\Vars)
  }
  \caption{Proof rule \ruleCTLInit}
  \label{fig-ctl-proof-rule-init}
\end{figure}
\end{minipage} %
\hspace{1 em}
\begin{minipage}{.48\textwidth} %
\begin{figure}[H]
  \makeFramedRule{}{
    p(\Vars) \limp \exists \Vars': \relnext(\Vars,\Vars') \land q(\Vars')
  }{
    (p(v), \relnext(v,v')) \ctlsat EX~q(\Vars)
  }
  \caption{Proof rule \ruleCTLEX}
  \label{fig-ctl-proof-rule-EX}
\end{figure}
\end{minipage}
\vspace{2 em}\\
The proof rules \ruleCTLEX (see Figure~\ref{fig-ctl-proof-rule-EX}),
\ruleCTLEG (see Figure~\ref{fig-ctl-proof-rule-EG}), and  \ruleCTLEU
(see Figure~\ref{fig-ctl-proof-rule-EU}) are applied for generating
Horn constraints when the CTL satisfaction problem has a basic CTL
formula with existential path operator. \\
\begin{minipage}{.4\textwidth} 
\begin{figure}[H]
  \makeFramedRuleSS{Find an assertion $\Inv(v)$ such that:}{
    p(\Vars) & \limp & \Inv(\Vars)\\
    \Inv(\Vars) & \limp & \exists \Vars': \relnext(\Vars,\Vars') \land \Inv(\Vars')\\
    \Inv(\Vars) & \limp & q(\Vars)
  }{
    (p(v), \relnext(v,v')) \ctlsat EG~q(\Vars)
  }
  \caption{Proof rule \ruleCTLEG }
  \label{fig-ctl-proof-rule-EG}
\end{figure}
\end{minipage} %
\hspace{1 em}
\begin{minipage}{.56\textwidth} %
\begin{figure}[H]
  \makeFramedRuleSS{Find assertions $\Inv(v)$ and $\rank(v,v')$ such that: }{
        p(\Vars) & \limp & \Inv(\Vars)\\
        \Inv(\Vars) \land \neg r(\Vars) &  \limp &  q(\Vars) \land \exists \Vars':  \relnext(\Vars,\Vars') ~\land \\[\jot]
                 & & \Inv(\Vars') \land \rank(\Vars,\Vars') \\[\jot]
       & \qsymWF(\rank)
  }{
    (p(v), \relnext(v,v')) \ctlsat EU(q(\Vars),r(\Vars))
  }
  \caption{Proof rule \ruleCTLEU}
  \label{fig-ctl-proof-rule-EU}
\end{figure}
\end{minipage}
\vspace{2 em}\\
%

Similarly, the proof rules \ruleCTLAX (see Figure~\ref{fig-ctl-proof-rule-AX}),
\ruleCTLAG (see Figure~\ref{fig-ctl-proof-rule-AG}), and  \ruleCTLAU
(see Figure~\ref{fig-ctl-proof-rule-AU}) are applied for generating
Horn constraints when the CTL satisfaction has a basic CTL formula with
universal path operator.

\begin{minipage}{.48\textwidth} 
\begin{figure}[H]
  \makeFramedRuleSS{}{
    p(\Vars) \land \relnext(\Vars,\Vars') & \limp & q(\Vars')
  }{
    (p(v), \relnext(v,v')) \ctlsat AX~q(\Vars)
  }
  \caption{Proof rule \ruleCTLAX}
  \label{fig-ctl-proof-rule-AX}
\end{figure}
\end{minipage} %
\hspace{1 em}
\begin{minipage}{.48\textwidth} %
\begin{figure}[H]
  \makeFramedRuleSS{Find an assertion $\Inv(v)$ such that:}{
    p(\Vars) & \limp & \Inv(\Vars)\\
    \Inv(\Vars) \land \relnext(\Vars,\Vars') & \limp & \Inv(\Vars')\\
    \Inv(\Vars) & \limp & q(\Vars)
  }{
    (p(v), \relnext(v,v')) \ctlsat AG~q(\Vars)
  }
  \caption{Proof rule \ruleCTLAG}
  \label{fig-ctl-proof-rule-AG}
\end{figure}
\end{minipage}
\vspace{2 em}\\
\begin{figure}[h]
  \makeFramedRule{Find assertions $\Inv(v)$ and $\rank(v,v')$ such that: }{
    p(\Vars) \limp \Inv(\Vars)\\
    \Inv(\Vars) \land \neg r(\Vars) \land \relnext(\Vars,\Vars') \limp
    \begin{array}[t]{l} 
      q(\Vars) \land \Inv(\Vars') \land \rank(\Vars,\Vars')
    \end{array}\\
    \qsymWF(\rank).
  }{
    (p(v), \relnext(v,v')) \ctlsat AU(q(\Vars), r(\Vars))
  }
  \caption{Proof rule \ruleCTLAU}
  \label{fig-ctl-proof-rule-AU}
\end{figure}

Our proof system is not exhaustive in terms of having proof rules for
all kinds of basic CTL formulas.
However, we utilize equivalence between CTL formulas to generate
Horn constraints for basic CTL formulas whose proof rules are not
given in the proof system. 
For example, the equivalence between the formulas
$EU(\ltrue,q(v))$ and $EF(q(v))$ can be used to reduce the CTL
satisfaction problem $(p(v), \relnext(v,v')) \ctlsat EF(q(v)$ into
$(p(v), \relnext(v,v')) \ctlsat EU(true, q(v))$.


\section{Constraint generation}
\label{sec:cons-gen}

The contraint generation procedure performs a top-down, recursive
descent through the syntax tree of the given CTL formula.
At each level of recursion, the procedure takes as input a CTL
satisfaction $(p(v),\relnext(v,v')) \ctlsat \varphi$, where $\varphi$
is a CTL formula and assertions $p(v)$ and
$\relnext(v, v')$ describe a set of states and a transition
relation, respectively.
The constraint generation procedure applies proof rules from the proof
system presented in the previous section to recursively decompose
complex satisfactions and eventually generate forall-exists quantified
Horn constraints with well-foundedness conditions.
Before starting the actual constraint generation, the procedure
recursively rewrites the input satisfaction of a given CTL formula with arbitrary
structure into a set of satisfactions of simple CTL formulas where
each simple formula is either a basic CTL state formula or an
assertion over the background theory.
The procedure then takes each satisfaction involving simple formula,
introduces auxiliary predicates and generates a sequence of
forall-exists quantified Horn constraints and well-foundedness
constraints (when needed) over these predicates.

\paragraph*{Complexity and Correctness}
The procedure performs a single top-down descent through the syntax tree
of the given CTL formula $\varphi$.
The run time for constraints generation, and hence the size of the
generated constraints, is linear in the size of $\varphi$.
Finding a solution for the generated Horn constraints is undecidable in
general.
In practice however, our solving algorithm \algEHSF often succeeds in
finding a solution (see Section~\ref{sec:eval}).
We formalize the correctness of the constraint generation procedure in
the following theorem.
\begin{theorem}
For a given program $\program$ with $\init(v)$ and $\nextRel(v, v')$ over $v$ and
a CTL formula $\varphi$ the Horn constraints generated from
$(p(v),\relnext(v,v')) \ctlsat \varphi$ are satisfiable if and only
if $\program \models \varphi$.
\end{theorem}
The proof can be found in~\cite{KestenTCS95}.

\paragraph*{Example}
Let us consider the program given in
Figure~\ref{fig-ctl-example-code}. 
It contains the variable \emph{rho} which is assigned a non-deterministic
value at Line~2.
This assignment results in the program control to move
non-deterministically following the evaluation of the condition at
Line~4.
It is common to verify such programs with respect to various CTL
properties as the non-determinism results in different computation
paths of the program. 
Now, we would like to verify the program with respect to the CTL property
$\mathit{AG(EF~(WItemsNum \geq 1))}$, i.e., from every reachable 
state of the program, there exists a path to a state where
\emph{WItemsNum} has a positive integer value. 
\begin{figure}[h]
  \begin{lstlisting}[language=C]
      int main () {
1:       while(1) {
2:         while(1) { 
              rho = nondet();
3:           if (WItemsNum<=5) { 
4:             if (rho>0) break; }
5:           WItemsNum++;
6:         } 
7:         while(1) { 
8:           if (!(WItemsNum>2)) break;
9:           WItemsNum--;
10:        }
11:      }
12:    }
  \end{lstlisting}
  \caption{An example program}
  \label{fig-ctl-example-code}
\end{figure}

We can make the following observations about the program.
The value of the variable \emph{WItemsNum} is not set initially. 
Therefore, the property is checked for any arbitrary initial value of
\emph{WItemsNum}.
The verification problem is more interesting for the case when
\emph{WItemsNum}  has a non-positive integer value.
\begin{figure}[h]
\begin{equation*}
  \begin{array}[t]{@{}rl@{\;}l@{\;}}
    \Vars & =& (w,pc)  \\[\jot]
    \init(\Vars) & = &(pc=1) \\[\jot]
    \relnext(v,v') & = &(pc=\loc_1 \land \pc'=\loc_2 \land w'=w ~\lor pc=\loc_2 \land \pc'=\loc_3 \land w'=w ~\lor \\[\jot]
          & & pc=\loc_3 \land w \leq 5 \land \pc'=\loc_4 \land w'=w ~\lor pc=\loc_3 \land w >5 \land \pc'=\loc_5 \land w'=w ~\lor \\[\jot]
          & & pc=\loc_4 \land \pc'=\loc_5 \land w'=w ~\lor pc=\loc_4 \land \pc'=\loc_7 \land w'=w ~\lor \\[\jot]
          & &  pc=\loc_5 \land \pc'=\loc_6 \land w'=w+1 ~\lor  pc=\loc_6 \land \pc'=\loc_3 \land w'=w ~\lor \\[\jot]
          & & pc=\loc_7 \land \pc'=\loc_8 \land w'=w ~\lor pc=\loc_8 \land w \leq 2 \land \pc'=\loc_{11} \land w'=w ~\lor \\[\jot]
          & & pc=\loc_8 \land w > 2 \land \pc'=\loc_9 \land w'=w ~\lor pc=\loc_9 \land \pc'=\loc_{10} \land w'=w-1 ~\lor \\[\jot]
          & & pc=\loc_{10} \land \pc'=\loc_8 \land w'=w ~\lor pc=\loc_{11} \land \pc'=\loc_3 \land w'=w)
  \end{array}
\end{equation*}
  \caption{Transition system for the example program}
  \label{fig-TS-example}
\end{figure}
This is because depending on how the variable \emph{rho} is
instantiated at Line~2, we may get a path that will not reach a state where
\emph{WItemsNum} gets a positive integer value.
For example, if we assume \emph{WItemsNum} has the value 0 initially
and \emph{WItemsNum} is instantiated to the value 1, the program
control swings between the two internal loops by keeping the value of
\emph{WItemsNum}  the same.
This resulting path will not reach the state with \emph{WItemsNum
  $\geq$ 1}. 
However, if \emph{rho} is assigned a non-positive value, no matter
what the value of \emph{rho} is initially, it will eventually reach a
value greater than 5 before exiting the first nested loop.
Such a path will eventually reach the state with \emph{WItemsNum
  $\geq$ 1} and hence the program satisfies the CTL property
$AG(EF~(WItemsNum \geq 1))$.

Our method abstracts away from the concrete syntax of a programming
language by modeling a program as a transition system.
The transition system for the program is given in Figure~\ref{fig-TS-example}.
In the tuple of variables $v$, the variable $w$ corresponds to the
program variable \emph{WItemsNum} and $pc$ is the program counter
variable.
The problem of verifying the program with respect to the given property
amounts to checking if $(\init(v), \relnext(v,v'))$ satisfies
$AG(EF(w \geq 1))$, i.e., if the satisfaction $(\init(v), \relnext(v,
v')) \ctlsat AG(EF(w \geq 1))$ holds.
Our method first generates a set of Horn constraint corresponding to
the verification problem by applying the proof system.

We start constraint generation by considering the nesting structure of
$AG(EF(w \geq 1))$.
Due to the fact that $AG(EF(w \geq 1))$ has $AG$ as the outermost operator, we apply $\ruleCTLDecompUni$
from Figure~\ref{fig-ctl-proof-rule-decompUni} to split the original
satisfaction $(\init(v), \relnext(v, v')) \ctlsat AG(EF(w \geq 1))$ into
a reduced satisfaction $(\init(v), \relnext(v, v')) \ctlsat AG(p_1(v))$ and
a new satisfaction $(p_1(v), \relnext(v, v')) \ctlsat EF(w \geq 1)$. 
We need to solve for the auxiliary assertion $p_1(v)$ satisfying both of
the satisfactions.

The assertion $p_1(v)$ corresponds to a set of program states that
needs to be discovered from the initial state. 
This is represented by the new satisfaction $(\init(v), \relnext(v, v'))
\ctlsat AG(p_1(v))$ which is reduced directly to a set of Horn
constraints by applying $\ruleCTLAG$
from Figure~\ref{fig-ctl-proof-rule-AG}.
This set of Horn constraints is over an auxiliary
predicate~$\mathit{inv}_1(v)$ and given below.
\begin{equation*}
  \begin{array}[t]{@{}l@{}}
    \mathit{init}(v) \limp \mathit{inv}_1(v),\\[\jot]
    \mathit{inv}_1(v) \land \mathit{next}(v, v') \limp \mathit{inv}_1(v'),\\[\jot]
    \mathit{inv}_1(v) \limp p_1(v).
  \end{array}
\end{equation*}
%

The formula $EF(w \geq 1)$, which was nested in the original formula
$AG(EF(w \geq 1))$, must also be satisfied from the set of
states represented by $p_1(v)$.
This is represented by the new satisfaction $(p_1(v), \relnext(v,
v')) \ctlsat EF(w \geq 1)$.
Such new satisfactions may not lead directly to Horn constraints
generation and may require further reduction into simpler
satisfactions.  
Since $EF(w \geq 1)$ has $EF$ as the outermost operator, we apply
again $\ruleCTLDecompUni$ from
Figure~\ref{fig-ctl-proof-rule-decompUni} to split the satisfaction
$(\p_1(v), \relnext(v, v')) \ctlsat EF(w \geq 1)$ into a reduced satisfaction
$(\p_1(v), \relnext(v, v')) \ctlsat EF(p_2(v))$ and a new
satisfaction $(p_2(v), \relnext(v, v')) \ctlsat w \geq 1$. 
Here also, we need to solve for the auxiliary assertion $p_2(v)$ satisfying both of
the satisfactions.

The equivalence between the formulas $EF(p_2 (v))$ and $EU(\ltrue, p_2
(v))$ is used to reduce the CTL satisfaction problem $(\p_1(v),
\relnext(v, v')) \ctlsat EF(p_2(v))$ into $(p_1(v), \relnext(v,v'))
\ctlsat EU(true, p_2(v))$. 
The corresponding set of Horn constraints are generated by applying
$\ruleCTLEU$ from Figure~\ref{fig-ctl-proof-rule-EU}. 
Due to the existential path quantifier in $EU(true, p_2(v))$, we
obtain clauses that contain existential quantification.
We deal with the eventuality by imposing a well-foundedness condition.
This set of Horn constraints is over the auxiliary
assertions $\inv_2(v)$ and $\rank(v, v')$, and it is given below.
\begin{equation*}
  \begin{array}[t]{@{}l@{}}
    p_1(v) \limp \mathit{inv}_2(v),\\[\jot]
    \mathit{inv}_2(v) \land \neg p_2(v) \limp 
    \exists v':  \mathit{next}(v, v') \land \inv_2(v') \land \mathit{rank}(v, v'),
    \\[\jot]
    \qsymWF(\rank)
  \end{array}
\end{equation*}

Coming to the new satisfaction $(p_2(v), \relnext(v, v')) \ctlsat
w \geq 1$, we see that its formula $w \geq 1$ is an assertion with no
temporal operators. 
Since no further decomposition is possible, we apply $\ruleCTLInit$ from
Figure~\ref{fig-ctl-proof-rule-init} to generate directly the
clause:
\begin{equation*}
    p_2(v) \limp w \geq 1
\end{equation*}

As the original satisfaction $(\init(v), \relnext(v, v')) \ctlsat
AG(EF(w \geq 1))$ is reduced into the satisfactions 
$(\init(v), \relnext(v, v')) \ctlsat AG(p_1(v))$, 
$(p_1(v), \relnext(v,v')) \ctlsat EF(p_2(v))$ and 
$(p_2(v), \relnext(v, v')) \ctlsat w \geq 1$, 
the constraints for the original satisfaction will be the union of the
constraints for each of the decomposed satisfactions.
The Horn constraints are over the auxiliary assertions $p_1(v)$,
$\inv_1(v)$, $p_2(v)$, $\inv_2(v)$ and $\rank(v, v')$, and they are
given below. 
\begin{equation*}
  \begin{array}[t]{@{}l@{}}
    \mathit{init}(v) \limp \mathit{inv}_1(v),\\[\jot]
    \mathit{inv}_1(v) \land \mathit{next}(v, v') \limp \mathit{inv}_1(v'),\\[\jot]
    \mathit{inv}_1(v) \limp p_1(v),\\[\jot]
    p_1(v) \limp \mathit{inv}_2(v),\\[\jot]
    \mathit{inv}_2(v) \land \neg p_2(v) \limp 
    \exists v':  \mathit{next}(v, v') \land \inv_2(v') \land \mathit{rank}(v, v'),
    \\[\jot]
    \qsymWF(\rank)  \\[\jot]
    p_2(v) \limp w \geq 1
  \end{array}
\end{equation*}
This will be the final output of our Horn constraint generation
procedure.


\section{Evaluation}
\label{sec:eval}

We evaluate our method of CTL verification by applying the
implementation of the \Tool solver on a set of industrial benchmarks
from~\cite[Figure~7]{CookPLDI13}. 
These benchmarks consists of seven programs:
\texttt{Windows OS fragment 1}, \texttt{Windows OS fragment 2},
\texttt{Windows OS fragment 3}, \texttt{Windows OS fragment 4},
\texttt{Windows OS fragment 5}, \texttt{PostgreSQL pgarch} and
\texttt{Software Updates}. 
For each of these programs, four slightly different versions are
considered for evaluation.
In general, the four versions of a given program are the same
in terms of the main logic of the program and what the program does,
but they may differ on the value assigned to a particular variable or
the condition for exiting a loop, etc.
This gives us a set of 28 programs.
Each such program $P$ is provided with a CTL property $\varphi$, and
there are two verification tasks associated with it: $P \ctlsat
\varphi$ and $P \ctlsat \neg\varphi$.
The existence of a proof for a property $\varphi$ for $P$ implies that $\neg\varphi$
is violated by the same program $P$, and similarly, a proof for
$\neg\varphi$ for $P$ implies that $\varphi$ is violated by $P$.
However, it may also be the case that neither $P \not\ctlsat \varphi$ nor $P \not\ctlsat
\neg\varphi$ hold.

\paragraph{Templates: }
As discussed in Section~\ref{subsec-ehsf}, \algEHSF requires the
template functions to be provided by the user for relations with
existentially quantified variables.    
For the application of CTL verification, which is the main topic of
interest in the paper, we claim that the transition relation
$\relnext(v,v')$ can be used as a template by adding constraints at
each location of non-determinism.
There are two kinds of constraints that can be added depending on the
two types of possible non-determinism in $\relnext(v,v')$.
\begin{itemize} 
\item \textbf{non-deterministic guards: } this is the case when
  $\relnext(v,v')$ has a set of more than one disjuncts with the
  same guard, i.e., there can be more than one enabled moves from a
  certain state of the program. 
  For each such set, we introduce a fresh case-splitting variable and
  we strengthen the guard of each disjunct by adding a distinct
  constraint on the fresh variable. 
  For example, if the set has $n$ disjuncts and $B$ is a fresh
  variable, we add the constraint $B=i$ for each disjunct $i$ where
  $1 \leq i \leq n$.
  To reason about existentially quantified queries, then it will
  suffices to instantiate $B$ to one of the values in the range
  $1\dots n$.
  Such reasoning is done by the \algEHSF solver. 
\item \textbf{non-deterministic assignments:} this is the case when
  $\relnext(v,v')$ has a disjunct in which some $w'$, which is a
  subset of $v'$, is left unconstrained in the disjunct.
  In such case, we strengthen the disjunct by adding the constraint
  $x'=T_x*v+t_x $ as conjunct for each variable $x'$ in $w'$.
  Solving for $T_x$ and $t_x$ is done by the \algEHSF solver. 
\end{itemize}
In our CTL verification examples, both non-deterministic guards and
assignments are explicitly marked in the original
benchmark programs using names~$\mathtt{rho1}, \mathtt{rho2},$ etc.
We apply the techniques discussed above to generate templates from the
transition relation of each program. 
In these examples, linear templates are sufficiently expressive. 
For dealing with well-foundedness we use linear ranking functions. 
\begin{table}[h]
\centering
\begin{tabular}{@{}|c|c|c|r|c|r|}
  \hline
  Program $P$ & Property $\varphi$ & \cc{$P \ctlsat \varphi$} & \cc{$P \ctlsat~\neg\varphi$} \\
  \cline{3-4} \cline{5-6}
  & & Result & Time(s) & Result & Time(s) \\
  \hline 
  \multirow{4}{*}{\parbox{0.3\linewidth}{\centering \texttt{Windows OS fragment 1 (29 LOC)}}}
              & $AG(p \limp AF q)$    & \ipYes & 0.3 &  \ipNo   & 0.3  \\
              & $EF(p \land EG q)$     & \ipYes & 0.3 & \ipNo    & 0.3 \\
              & $AG(p \limp EF q)$    & \ipYes & 0.3 &  \ipNo   & 0.3 \\
              & $EF(p \land AG q)$    & \ipYes & 0.3 &\ipNo    & 0.3 \\
  \hline 
  \multirow{4}{*}{\parbox{0.3\linewidth}{\centering \texttt{Windows OS fragment 2 (58 LOC)}}}
              & $AG(p \limp AF q)$    & \ipYes & 0.4 &  \ipNo & 0.3 \\
              & $EF(p \land EG q)$     & \ipYes & 0.4 & \ipNo & 0.3  \\
              & $AG(p \limp EF q)$    & \ipYes & 0.4& \ipNo & 0.3 \\
              & $EF(p \land AG q)$    & \ipYes & 0.4& \ipNo & 0.3 \\
  \hline 
  \multirow{4}{*}{\parbox{0.3\linewidth}{\centering \texttt{Windows OS fragment 3 (370 LOC)}}}
              & $AG(p\limp AF q)$   & \ipYes  & 0.6  & \ipNo  & 1.2 \\
              & $EF(p\land EG q)$     & \ipYes  & 9.4 & \ipNo &  0.5 \\
              & $AG(p\limp EF q)$    & \ipYes  & 0.7  & \ipNo  & 0.8   \\
              & $EF(p\land AG q)$    & \ipYes & 0.9  & \ipNo  & 1.1  \\
  \hline 
  \multirow{4}{*}{\parbox{0.3\linewidth}{\centering \texttt{Windows OS fragment 4 (380 LOC)}}}    
              & $AF p \lor AF q$    & \ipYes & 5.7   & \ipNo    & 5.2      \\
              & $EG p \land EG q$  & \ipYes & 0.3      & \ipNo    & 1.0 \\
              & $EF p \land EF q$   & \ipYes & 5.0  & \ipNo    & 0.3    \\
              & $AG p \lor AG q$   & \ipYes  & 0.3     & \ipNo    & 6.4  \\
  \hline 
\multirow{4}{*}{\parbox{0.3\linewidth}{\centering \texttt{Windows OS fragment 5 (43 LOC)}}}
              & $AG(AF p)$  & \ipYes       & 0.3 & \ipNo & 0.3 \\
              & $EF(EG p)$   & \ipYes       & 0.3 & \ipNo & 0.3 \\ 
              & $AG(EF p)$   & \ipYes       & 0.3 & \ipNo & 0.3 \\ 
              & $EF(AG p)$   & \ipYes       & 0.3 & \ipNo & 0.3 \\ 
  \hline 
  \multirow{4}{*}{\parbox{0.25\linewidth}{\centering \texttt{PostgreSQL pgarch (70 LOC)}}}
              & $AG(AF p)$  & \ipYes  & 0.4 & \ipNo   & 0.3 \\
              & $EF(EG p)$   & \ipYes  & 0.3 & \ipNo   & 0.4 \\ 
              & $AG(EF p)$  & \ipYes  & 0.3 & \ipNo   & 0.3 \\ 
              & $EF(AG p)$  & \ipYes  & 0.3 & \ipNo   & 0.3 \\ 
  \hline 
  \multirow{4}{*}{\parbox{0.25\linewidth}{\centering \texttt{Software Updates (35 LOC)}}}
              & $p \limp EF q$  & \ipYes & 0.6 & \ipNo & 0.2 \\ 
              & $p \land EG q$  & \ipNo  & 0.3 & \ipNo & 0.4 \\
              & $p \limp AF q$  & \ipNo  & 0.2 & \ipNo & 0.2 \\
              & $p \land AG q$ & \ipNo  & 0.3 & \ipNo & 0.3 \\
  \hline
\end{tabular}

\caption{CTL verification on industrial benchmarks}
\label{table-ehsf-eval}
\end{table}


We report the results in Table~\ref{table-ehsf-eval}.
For each program in Column~1, we report the shape of the property
$\varphi$ in Column~2. 
The variables $p$ and $q$ in Column~2 range over the theory of
quantifier-free linear integer arithmetic. 
The result as well as the time it took the \algEHSF engine to prove the property
$\varphi$ is given in Columns~3~and~4, and similarly, the result as well
as the time it took the engine to discover a counterexample for the
negated property $\neg\varphi$ is given in Columns~5~and~6. 
The symbol \ipYes marks the cases where \algEHSF was able to find a
solution, i.e., a proof that the CTL property $\varphi$ is valid, and 
the symbol \ipNo~marks the cases where \algEHSF was able to find a
counter-example, i.e., a proof that the negated CTL property
$\neg\varphi$ is not valid.
The number of LOC of each program is also given in Column~1.

The \algEHSF engine is able to find proofs that the CTL property $\varphi$ is
valid (and the negated CTL property $\neg\varphi$ is not
valid) for all of the programs except the last three programs.
For the last three versions of \texttt{Software Updates}, not only the
negated CTL property $\neg\varphi$ but also the CTL property $\varphi$
is not valid.
This was because $\varphi$ was satisfied only for some initial states.
The method takes a total time of 52 seconds to complete the
verifications tasks.  
\begin{table}[h]
\centering
\begin{tabular}{@{}|c|c|c|r|c|r|}
  \hline
  Program $P$ & Property $\varphi$ & \cc{$P \ctlsat \varphi$} & \cc{$P \ctlsat~\neg\varphi$} \\
  \cline{3-4} \cline{5-6}
  & & \algEHSF & Cook et al. & \algEHSF & Cook et al. \\
  \hline 
  \multirow{4}{*}{\parbox{0.3\linewidth}{\centering \texttt{Windows OS fragment 1 (29 LOC)}}}
              & $AG(p \limp AF q)$    & 0.3 & 1.0 & 0.3 & 1.4 \\
              & $EF(p \land EG q)$     & 0.3 &  0.1   & 0.3 & 0.7\\
              & $AG(p \limp EF q)$    & 0.3 &   0.1  & 0.3 & 0.1\\
              & $EF(p \land AG q)$    & 0.3 & 0.1   & 0.3 & 0.1\\
  \hline 
  \multirow{2}{*}{\parbox{0.3\linewidth}{\centering \texttt{Windows OS fragment 2 (58 LOC)}}}
              & $EF(p \land EG q)$     & 0.4 & 1.0 & 0.3  & 1.2 \\
              & $EF(p \land AG q)$    & 0.4& 0.8 & 0.3 & 0.2\\
  \hline 
  \multirow{4}{*}{\parbox{0.3\linewidth}{\centering \texttt{Windows OS fragment 3 (370 LOC)}}}
              & $AG(p\limp AF q)$    & 0.6  & 5.9  & 1.2 & 6.2 \\
              & $EF(p\land EG q)$      & 9.4 & 2.3 &  0.5 & 6.0\\
              & $AG(p\limp EF q)$     & 0.7  & 6.8  & 0.8   & 3.4\\
              & $EF(p\land AG q)$    & 0.9  & 4.7  & 1.1  & 3.1\\
  \hline 
  \multirow{4}{*}{\parbox{0.3\linewidth}{\centering \texttt{Windows OS fragment 4 (380 LOC)}}}    
              & $AF p \lor AF q$    & 5.7   &  18.5   & 5.2 & 13.9\\
              & $EG p \land EG q$  & 0.3  &  13.5   & 1.0 & 14.2\\
              & $EF p \land EF q$   & 5.0  &  14.7 & 0.3  & 4.8\\
              & $AG p \lor AG q$    & 0.3     &  8.0   & 6.4  & 3.7\\
  \hline 
\multirow{4}{*}{\parbox{0.3\linewidth}{\centering \texttt{Windows OS fragment 5 (43 LOC)}}}
              & $AG(AF p)$        & 0.3 & 1.0 & 0.3 & 0.2\\
              & $EF(EG p)$         & 0.3 & 0.1 & 0.3 & 0.0\\ 
              & $AG(EF p)$         & 0.3 & 1.0 & 0.3 & 0.0\\ 
              & $EF(AG p)$         & 0.3 & 0.1 & 0.3 & 0.1\\ 
  \hline 
  \multirow{4}{*}{\parbox{0.25\linewidth}{\centering \texttt{PostgreSQL pgarch (70 LOC)}}}
              & $AG(AF p)$   & 0.4 & 2.0   & 0.3 & 1.3\\
              & $EF(EG p)$    & 0.3 &  0.1  & 0.4 & 0.1\\ 
              & $AG(EF p)$   & 0.3 &  2.0  & 0.3 & 0.0\\ 
              & $EF(AG p)$   & 0.3 &  2.0  & 0.3 & 2.4\\ 
  \hline
\end{tabular}

\caption{Comparison of our results with Cook~\cite[Figure~11]{Cook2014FTR}}
\label{table-ehsf-comp}
\end{table}


%
Our method also compares favourably with state-of-art automated CTL 
verification methods. 
We present in Table~\ref{table-ehsf-comp} the comparison between the
our solving algorithm \algEHSF and a CTL verification method from
Cook~\cite{Cook2014FTR}.
Here also, we use the programs from Table~\ref{table-ehsf-eval},
however, for the sake of focusing on the comparison, we exclude
programs for which the two methods have different outcomes.
For each program in Column~1, we report the shape of the property in
Column~2.
The time it takes \algEHSF to prove the property $\varphi$ is given in
Column~3, and the corresponding time for Cook is given in Column~4.
Similarly, the time it takes \algEHSF to discover a counterexample for
the negated property $\neg \varphi$ is given in
Column~5, and the corresponding time for Cook is given in Column~6.

From the result, we can see that while \algEHSF takes a total of 48
seconds to finish the task, Cook takes a total of
149 seconds. 
This amounts to an approximate reduction of 70\%. 
There are a few cases where \algEHSF takes longer than
Cook.
We suspect that a more efficient modeling of the original c program as
a transition system can help our method a lot. 
The presence of many temporary program variables in the transition
relation which are not involved in any computation of the program can
affect the performance of our method.



\section{Related work}
\label{sec:rel-work}

Verification of properties specified in temporal logics such as CTL
has been extensively explored for finite-state
systems\cite{Kupferman2000, Burch1990, Clarke02treelikeCEX,
  Clarke1983AVF}. 
There has also been studies on the verification of CTL properties for
some restricted types of infinite-state systems.
Some examples are pushdown processes \cite{Song2013, Song2011, Walukiewicz2000},
pushdown games \cite{walukiewicz2001pushdown},
and parameterised systems \cite{Emerson1996}.
For such restricted systems, the standard procedure is to abstract the
infinite-state system model into finite-state model and
apply the known methods for finite-state systems. 
But existing abstraction methods usually do not allow reliable
verification of CTL properties where alternation between universal and
existential modal operators is common. 
Many methods of proving CTL properties with only universal path
quantifiers are known\cite{Chaki2005, Cook2012TPV}.
 There also a few methods mainly focused on proving branching-time
properties with only existential path quantifiers.
One example is the tool Yasm \cite{GurfinkelWC06} which implements a
proof procedure aimed primarily at the non-nested existential subset of
CTL.
There are also known techniques for proving program
termination~\cite{Bradley05polyrankingfor, TerminatorPLDI06}
(resp. non-termination~\cite{TNTPOPL08}) which is equivalent with
proving the CTL formula $AF~false$ (resp. $EG~true$).

Banda et al.~\cite{Banda2010CAS} proposed a CTL verification approach
for infinite state reactive systems based on CLP and abstraction of a
CTL semantic function. 
An automatic proof method that supports both universal and existential
branching-time modal operators for (possibly infinite-state) programs
is proposed in by Cook et al.~\cite{CookPLDI13}. 
Cook's approach is based on reducing existential reasoning to universal
reasoning when an appropriate restriction is placed on the the
state-space of the system. 
While this approach comes close to our approach, the refinement
procedure for state-space restrictions may make
incorrect choices early during the iterative proof search.
These choices may limit the choices available later in the search
leading to  failed proof attempts in some cases.

\section{Conclusion}
\label{sec:concl}

In this paper, we proposed a method for verifying CTL properties with
respect to a (possibly infinite-space) program. 
The method takes a transition system that models the input program and
a CTL formula specifying the property to prove as inputs.
It first applies proof rules from its proof system to generate a set
of forall-exists quantified Horn constraints and well-foundedness
constraints. 
Then, it applies the solving algorithms \algEHSF to solve the
generated set of Horn constraints.
The defining feature of this approach is the separation of concerns
between the encoding and the solving of the verification problem.
Although our method is based on generic Horn constraint solving
engine, it is able to outperform state-of-art methods 
specialised for CTL verification.
We also demonstrate the practical applicability of the approach by
presenting an experimental evaluation using examples from the
PostgreSQL database server, the SoftUpdates patch system, the Windows
OS kernel.


\newpage
\bibliographystyle{eptcs}
\bibliography{biblio}

\begin{thebibliography}{10}
\providecommand{\bibitemdeclare}[2]{}
\providecommand{\surnamestart}{}
\providecommand{\surnameend}{}
\providecommand{\urlprefix}{Available at }
\providecommand{\url}[1]{\texttt{#1}}
\providecommand{\href}[2]{\texttt{#2}}
\providecommand{\urlalt}[2]{\href{#1}{#2}}
\providecommand{\doi}[1]{doi:\urlalt{http://dx.doi.org/#1}{#1}}
\providecommand{\bibinfo}[2]{#2}

\bibitemdeclare{inproceedings}{Banda2010CAS}
\bibitem{Banda2010CAS}
\bibinfo{author}{Gourinath \surnamestart Banda\surnameend} \&
  \bibinfo{author}{John~P. \surnamestart Gallagher\surnameend}
  (\bibinfo{year}{2010}): \emph{\bibinfo{title}{Constraint-based Abstract
  Semantics for Temporal Logic: A Direct Approach to Design and
  Implementation}}.
\newblock In: {\sl \bibinfo{booktitle}{Proceedings of the 16th International
  Conference on Logic for Programming, Artificial Intelligence, and
  Reasoning}}, \bibinfo{series}{LPAR'10}, \doi{10.1007/978-3-642-17511-4\_3}.

\bibitemdeclare{inproceedings}{ehsf}
\bibitem{ehsf}
\bibinfo{author}{Tewodros~A. \surnamestart Beyene\surnameend},
  \bibinfo{author}{Corneliu \surnamestart Popeea\surnameend} \&
  \bibinfo{author}{Andrey \surnamestart Rybalchenko\surnameend}
  (\bibinfo{year}{2013}): \emph{\bibinfo{title}{Solving Existentially
  Quantified Horn Clauses}}.
\newblock In: {\sl \bibinfo{booktitle}{Proceedings of the 25th International
  Conference on Computer Aided Verification}}, \bibinfo{series}{CAV'13},
  \doi{10.1007/978-3-642-39799-8\_61}.

\bibitemdeclare{article}{Bradley05polyrankingfor}
\bibitem{Bradley05polyrankingfor}
\bibinfo{author}{Aaron~R. \surnamestart Bradley\surnameend},
  \bibinfo{author}{Zohar \surnamestart Manna\surnameend} \&
  \bibinfo{author}{Henny~B. \surnamestart Sipma\surnameend}
  (\bibinfo{year}{2005}): \emph{\bibinfo{title}{Polyranking for Polynomial
  Loops}}.
\newblock {\sl \bibinfo{journal}{Automata, Languages and Programming}}, pp.
  \bibinfo{pages}{1349--1361}.

\bibitemdeclare{inproceedings}{Burch1990}
\bibitem{Burch1990}
\bibinfo{author}{J.R. \surnamestart Burch\surnameend}, \bibinfo{author}{E.M.
  \surnamestart Clarke\surnameend}, \bibinfo{author}{K.L. \surnamestart
  McMillan\surnameend}, \bibinfo{author}{D.L. \surnamestart Dill\surnameend} \&
  \bibinfo{author}{L.J. \surnamestart Hwang\surnameend} (\bibinfo{year}{1990}):
  \emph{\bibinfo{title}{Symbolic model checking: 10$^{20}$ states and beyond}}.
\newblock \doi{10.1016/0890-5401(92)90017-A}.

\bibitemdeclare{inproceedings}{Chaki2005}
\bibitem{Chaki2005}
\bibinfo{author}{Sagar \surnamestart Chaki\surnameend},
  \bibinfo{author}{Edmund~M. \surnamestart Clarke\surnameend},
  \bibinfo{author}{Orna \surnamestart Grumberg\surnameend},
  \bibinfo{author}{Joël \surnamestart Ouaknine\surnameend},
  \bibinfo{author}{Natasha \surnamestart Sharygina\surnameend},
  \bibinfo{author}{Tayssir \surnamestart Touili\surnameend} \&
  \bibinfo{author}{Helmut \surnamestart Veith\surnameend}
  (\bibinfo{year}{2005}): \emph{\bibinfo{title}{State/Event Software
  Verification for Branching-Time Specifications.}}
\newblock In: {\sl \bibinfo{booktitle}{IFM}}, \bibinfo{volume}{3771},
  \bibinfo{publisher}{Springer}, pp. \bibinfo{pages}{53--69},
  \doi{10.1007/11589976\_5}.

\bibitemdeclare{inproceedings}{Clarke1983AVF}
\bibitem{Clarke1983AVF}
\bibinfo{author}{E.~M. \surnamestart Clarke\surnameend}, \bibinfo{author}{E.~A.
  \surnamestart Emerson\surnameend} \& \bibinfo{author}{A.~P. \surnamestart
  Sistla\surnameend} (\bibinfo{year}{1983}): \emph{\bibinfo{title}{Automatic
  Verification of Finite State Concurrent System Using Temporal Logic
  Specifications: A Practical Approach}}.
\newblock In: {\sl \bibinfo{booktitle}{Proceedings of the 10th ACM
  SIGACT-SIGPLAN Symposium on Principles of Programming Languages}},
  \bibinfo{series}{POPL '83}, \doi{10.1145/567067.567080}.

\bibitemdeclare{inproceedings}{Clarke02treelikeCEX}
\bibitem{Clarke02treelikeCEX}
\bibinfo{author}{Edmund \surnamestart Clarke\surnameend}, \bibinfo{author}{Yuan
  \surnamestart Lu\surnameend}, \bibinfo{author}{Broadcom \surnamestart
  Com\surnameend}, \bibinfo{author}{Helmut \surnamestart Veith\surnameend} \&
  \bibinfo{author}{Somesh \surnamestart Jha\surnameend} (\bibinfo{year}{2002}):
  \emph{\bibinfo{title}{Tree-Like Counterexamples in Model Checking}}.
\newblock In: {\sl \bibinfo{booktitle}{In Proceedings of the 17 th Annual IEEE
  Symposium on Logic in Computer Science (LICS’02)}},
  \bibinfo{publisher}{IEEE Computer Society}, \doi{10.1109/LICS.2002.1029814}.

\bibitemdeclare{inproceedings}{Clarke1990TLM}
\bibitem{Clarke1990TLM}
\bibinfo{author}{Edmund~M. \surnamestart Clarke\surnameend}
  (\bibinfo{year}{1991}): \emph{\bibinfo{title}{Temporal Logic Model Checking:
  Two Techniques for Avoiding the State Explosion Problem}}.
\newblock \bibinfo{series}{CAV '90}.

\bibitemdeclare{inproceedings}{Cook2014FTR}
\bibitem{Cook2014FTR}
\bibinfo{author}{Byron \surnamestart Cook\surnameend}, \bibinfo{author}{Heidy
  \surnamestart Khlaaf\surnameend} \& \bibinfo{author}{Nir \surnamestart
  Piterman\surnameend} (\bibinfo{year}{2014}): \emph{\bibinfo{title}{Faster
  Temporal Reasoning for Infinite-State Programs}}.
\newblock \doi{10.1109/FMCAD.2014.6987598}.

\bibitemdeclare{inproceedings}{CookPLDI13}
\bibitem{CookPLDI13}
\bibinfo{author}{Byron \surnamestart Cook\surnameend} \& \bibinfo{author}{Eric
  \surnamestart Koskinen\surnameend} (\bibinfo{year}{2013}):
  \emph{\bibinfo{title}{Reasoning about Nondeterminism in Programs}}.
\newblock In: {\sl \bibinfo{booktitle}{PLDI}}, \doi{10.1145/2491956.2491969}.

\bibitemdeclare{article}{Cook2012TPV}
\bibitem{Cook2012TPV}
\bibinfo{author}{Byron \surnamestart Cook\surnameend}, \bibinfo{author}{Eric
  \surnamestart Koskinen\surnameend} \& \bibinfo{author}{Moshe \surnamestart
  Vardi\surnameend} (\bibinfo{year}{2012}): \emph{\bibinfo{title}{Temporal
  Property Verification As a Program Analysis Task}}.
\newblock {\sl \bibinfo{journal}{Form. Methods Syst. Des.}},
  \doi{10.1007/s10703-012-0153-5}.

\bibitemdeclare{inproceedings}{TerminatorPLDI06}
\bibitem{TerminatorPLDI06}
\bibinfo{author}{Byron \surnamestart Cook\surnameend}, \bibinfo{author}{Andreas
  \surnamestart Podelski\surnameend} \& \bibinfo{author}{Andrey \surnamestart
  Rybalchenko\surnameend} (\bibinfo{year}{2006}):
  \emph{\bibinfo{title}{Termination proofs for systems code}}.
\newblock In: {\sl \bibinfo{booktitle}{PLDI}}, \doi{10.1145/1133981.1134029}.

\bibitemdeclare{article}{Demri2010checkingctl}
\bibitem{Demri2010checkingctl}
\bibinfo{author}{St\'ephane \surnamestart Demri\surnameend},
  \bibinfo{author}{Alain \surnamestart Finkel\surnameend},
  \bibinfo{author}{Valentin~Goranko \surnamestart Govert\surnameend} \&
  \bibinfo{author}{Van \surnamestart Drimmelen\surnameend}
  (\bibinfo{year}{2010}): \emph{\bibinfo{title}{Model checking CTL* over flat
  Presburger counter systems}}.
\newblock {\sl \bibinfo{journal}{JANCL}}, \doi{10.3166/jancl.20.313-344}.

\bibitemdeclare{incollection}{Emerson1991TM}
\bibitem{Emerson1991TM}
\bibinfo{author}{E.~Allen \surnamestart Emerson\surnameend}
  (\bibinfo{year}{1990}): \emph{\bibinfo{title}{Handbook of Theoretical
  Computer Science (Vol. B)}}.
\newblock chapter \bibinfo{chapter}{Temporal and Modal Logic}.

\bibitemdeclare{inproceedings}{Emerson1996}
\bibitem{Emerson1996}
\bibinfo{author}{E.~Allen \surnamestart Emerson\surnameend} \&
  \bibinfo{author}{Kedar~S. \surnamestart Namjoshi\surnameend}
  (\bibinfo{year}{1996}): \emph{\bibinfo{title}{Automatic Verification of
  Parameterized Synchronous Systems (Extended Abstract)}}.
\newblock In: {\sl \bibinfo{booktitle}{Proceedings of the 8th International
  Conference on Computer Aided Verification}}, \bibinfo{series}{CAV '96},
  \doi{10.1007/3-540-61474-5\_60}.

\bibitemdeclare{inproceedings}{GrebenshchikovTACAS12}
\bibitem{GrebenshchikovTACAS12}
\bibinfo{author}{Sergey \surnamestart Grebenshchikov\surnameend},
  \bibinfo{author}{Ashutosh \surnamestart Gupta\surnameend},
  \bibinfo{author}{Nuno~P. \surnamestart Lopes\surnameend},
  \bibinfo{author}{Corneliu \surnamestart Popeea\surnameend} \&
  \bibinfo{author}{Andrey \surnamestart Rybalchenko\surnameend}
  (\bibinfo{year}{2012}): \emph{\bibinfo{title}{{HSF(C)}: A Software Verifier
  Based on {H}orn Clauses}}.
\newblock In: {\sl \bibinfo{booktitle}{TACAS}},
  \doi{10.1007/978-3-642-28756-5\_46}.

\bibitemdeclare{inproceedings}{TNTPOPL08}
\bibitem{TNTPOPL08}
\bibinfo{author}{Ashutosh \surnamestart Gupta\surnameend},
  \bibinfo{author}{Thomas~A. \surnamestart Henzinger\surnameend},
  \bibinfo{author}{Rupak \surnamestart Majumdar\surnameend},
  \bibinfo{author}{Andrey \surnamestart Rybalchenko\surnameend} \&
  \bibinfo{author}{Ru-Gang \surnamestart Xu\surnameend} (\bibinfo{year}{2008}):
  \emph{\bibinfo{title}{Proving non-termination}}.
\newblock In: {\sl \bibinfo{booktitle}{POPL}}, \doi{10.1145/1328438.1328459}.

\bibitemdeclare{inproceedings}{GurfinkelWC06}
\bibitem{GurfinkelWC06}
\bibinfo{author}{Arie \surnamestart Gurfinkel\surnameend},
  \bibinfo{author}{Ou~\surnamestart Wei\surnameend} \& \bibinfo{author}{Marsha
  \surnamestart Chechik\surnameend} (\bibinfo{year}{2006}):
  \emph{\bibinfo{title}{Yasm: A Software Model-Checker for Verification and
  Refutation.}}
\newblock In \bibinfo{editor}{Thomas \surnamestart Ball\surnameend} \&
  \bibinfo{editor}{Robert~B. \surnamestart Jones\surnameend}, editors: {\sl
  \bibinfo{booktitle}{CAV}}, \bibinfo{series}{Lecture Notes in Computer
  Science}, \bibinfo{publisher}{Springer}, pp. \bibinfo{pages}{170--174},
  \doi{10.1007/11817963\_18}.

\bibitemdeclare{inproceedings}{Hassan2012IIC}
\bibitem{Hassan2012IIC}
\bibinfo{author}{Zyad \surnamestart Hassan\surnameend},
  \bibinfo{author}{Aaron~R. \surnamestart Bradley\surnameend} \&
  \bibinfo{author}{Fabio \surnamestart Somenzi\surnameend}
  (\bibinfo{year}{2012}): \emph{\bibinfo{title}{Incremental, Inductive CTL
  Model Checking}}.
\newblock In: {\sl \bibinfo{booktitle}{Proceedings of the 24th International
  Conference on Computer Aided Verification}},
  \doi{10.1007/978-3-642-31424-7\_38}.

\bibitemdeclare{inproceedings}{muz}
\bibitem{muz}
\bibinfo{author}{Krystof \surnamestart Hoder\surnameend},
  \bibinfo{author}{Nikolaj \surnamestart Bj{\o}rner\surnameend} \&
  \bibinfo{author}{Leonardo \surnamestart de~Moura\surnameend}
  (\bibinfo{year}{2011}): \emph{\bibinfo{title}{{$\mu$Z}- An Efficient Engine
  for Fixed Points with Constraints}}.
\newblock In: {\sl \bibinfo{booktitle}{CAV}},
  \doi{10.1007/978-3-642-22110-1\_36}.

\bibitemdeclare{article}{KestenTCS95}
\bibitem{KestenTCS95}
\bibinfo{author}{Yonit \surnamestart Kesten\surnameend} \&
  \bibinfo{author}{Amir \surnamestart Pnueli\surnameend}
  (\bibinfo{year}{2005}): \emph{\bibinfo{title}{A compositional approach to
  {CTL}* verification}}.
\newblock {\sl \bibinfo{journal}{Theor. Comput. Sci.}}
  \bibinfo{volume}{331}(\bibinfo{number}{2-3}), pp. \bibinfo{pages}{397--428},
  \doi{10.1016/j.tcs.2004.09.023}.

\bibitemdeclare{article}{Kupferman2000}
\bibitem{Kupferman2000}
\bibinfo{author}{Orna \surnamestart Kupferman\surnameend},
  \bibinfo{author}{Moshe~Y. \surnamestart Vardi\surnameend} \&
  \bibinfo{author}{Pierre \surnamestart Wolper\surnameend}
  (\bibinfo{year}{2000}): \emph{\bibinfo{title}{An Automata-theoretic Approach
  to Branching-time Model Checking}}.
\newblock {\sl \bibinfo{journal}{J. ACM}}, \doi{10.1145/333979.333987}.

\bibitemdeclare{book}{Manna1992}
\bibitem{Manna1992}
\bibinfo{author}{Zohar \surnamestart Manna\surnameend} \& \bibinfo{author}{Amir
  \surnamestart Pnueli\surnameend} (\bibinfo{year}{1992}):
  \emph{\bibinfo{title}{The Temporal Logic of Reactive and Concurrent
  Systems}}.
\newblock \bibinfo{publisher}{Springer-Verlag New York, Inc.},
  \bibinfo{address}{New York, NY, USA}, \doi{10.1007/978-1-4612-0931-7}.

\bibitemdeclare{book}{McMillan1993SMC}
\bibitem{McMillan1993SMC}
\bibinfo{author}{Kenneth~L. \surnamestart McMillan\surnameend}
  (\bibinfo{year}{1993}): \emph{\bibinfo{title}{Symbolic Model Checking}}.
\newblock \doi{10.1007/978-1-4615-3190-6}.

\bibitemdeclare{article}{Penczek2002BMC}
\bibitem{Penczek2002BMC}
\bibinfo{author}{Wojciech \surnamestart Penczek\surnameend},
  \bibinfo{author}{Bozena \surnamestart Wozna\surnameend} \&
  \bibinfo{author}{Andrzej \surnamestart Zbrzezny\surnameend}
  (\bibinfo{year}{2002}): \emph{\bibinfo{title}{Bounded Model Checking for the
  Universal Fragment of CTL}}.
\newblock {\sl \bibinfo{journal}{Fundam. Inf.}}

\bibitemdeclare{inproceedings}{Pnueli1977}
\bibitem{Pnueli1977}
\bibinfo{author}{Amir \surnamestart Pnueli\surnameend} (\bibinfo{year}{1977}):
  \emph{\bibinfo{title}{The Temporal Logic of Programs}}.
\newblock In: {\sl \bibinfo{booktitle}{Proceedings of the 18th Annual Symposium
  on Foundations of Computer Science}}, \doi{10.1109/SFCS.1977.32}.

\bibitemdeclare{inproceedings}{Song2011}
\bibitem{Song2011}
\bibinfo{author}{Fu~\surnamestart Song\surnameend} \& \bibinfo{author}{Tayssir
  \surnamestart Touili\surnameend} (\bibinfo{year}{2011}):
  \emph{\bibinfo{title}{Efficient CTL model-checking for pushdown systems}}.
\newblock In: {\sl \bibinfo{booktitle}{In CONCUR}},
  \doi{10.1007/978-3-642-23217-6\_29}.

\bibitemdeclare{inproceedings}{Song2013}
\bibitem{Song2013}
\bibinfo{author}{Fu~\surnamestart Song\surnameend} \& \bibinfo{author}{Tayssir
  \surnamestart Touili\surnameend} (\bibinfo{year}{2013}):
  \emph{\bibinfo{title}{PoMMaDe: Pushdown Model-checking for Malware
  Detection}}.
\newblock In: {\sl \bibinfo{booktitle}{Proceedings of the 2013 9th Joint
  Meeting on Foundations of Software Engineering}}, \bibinfo{publisher}{ACM},
  \doi{10.1007/978-3-642-28756-5\_9}.

\bibitemdeclare{inproceedings}{Walukiewicz2000}
\bibitem{Walukiewicz2000}
\bibinfo{author}{Igor \surnamestart Walukiewicz\surnameend}
  (\bibinfo{year}{2000}): \emph{\bibinfo{title}{Model Checking CTL Properties
  of Pushdown Systems.}}
\newblock In: {\sl \bibinfo{booktitle}{FSTTCS}}, \bibinfo{series}{Lecture Notes
  in Computer Science}, \doi{10.1007/3-540-44450-5\_10}.

\bibitemdeclare{article}{walukiewicz2001pushdown}
\bibitem{walukiewicz2001pushdown}
\bibinfo{author}{Igor \surnamestart Walukiewicz\surnameend}
  (\bibinfo{year}{2001}): \emph{\bibinfo{title}{Pushdown processes: Games and
  model-checking}}.
\newblock {\sl \bibinfo{journal}{Information and computation}},
  \doi{10.1007/3-540-61474-5\_58}.

\end{thebibliography}
\end{document}